\documentclass[10pt,twocolumn,english,showpacs,preprintnumbers,amsmath,amssymb,prb,aps,preprint]{revtex4}

\usepackage[T1]{fontenc}
\usepackage[latin9]{inputenc}
\usepackage{textcomp}
\usepackage{amstext}
\usepackage{graphicx}

\makeatletter

\usepackage{epsfig}

\makeatother

\usepackage{babel}
\begin{document}

\title{Thermal control of graphene morphology: a signature of its intrinsic
surface tension}

\author{R. Ram\'{\i}rez and C. P. Herrero}

\affiliation{Instituto de Ciencia de Materiales de Madrid (ICMM), 
            Consejo Superior de Investigaciones Cient\'{\i}ficas (CSIC), 
            Campus de Cantoblanco, 28049 Madrid, Spain }

\begin{abstract}
The surface tension $\sigma$ of free-standing graphene is studied
by path-integral simulations as a function of the temperature and
the in-plane stress. Even if the applied stress vanishes, the membrane
displays a finite surface tension $\sigma$ due to the coupling between
the bending oscillations and the real area of the membrane. Zero-point
effects for $\sigma$ are significant below 100 K. Thermal cooling
drives the membrane from a planar to a wrinkled morphology. Upon heating
the change is reversible and shows hysteresis, in agreement to recent
experiments performed on supported graphene.
\end{abstract}


\maketitle

\section{Introduction}

Graphene, in contrast to the complex lipid bilayer membranes, is an
ideal system to understand physical effects of a two-dimensional (2D)
layer fluctuating in 3D space.\citep{roldan17,amorim16} Several experiments
have shown that the morphology of graphene can be reversibly changed
from a wrinkled configuration at low temperature to a planar one at
high temperature, typically in a range of 100-600 K.\citealp{bao09,hattab11,bao12,zhang_13,bai14,deng15,meng13}
This change has been qualitatively explained by the presence of an
underlying substrate. The mismatch between the expansion coefficients
of the substrate and graphene should produce mechanical stresses that
drive the morphology change. Wrinkles found in the experiments were
periodic and static, with amplitudes several orders of magnitude larger
than those arising from thermal fluctuations.\citep{bao09} 
Such planar-to-wrinkled
transition may, however, be a pure thermal effect and equally appear
in free-standing graphene, in the absence of any substrate. In fact,
for graphene under a small compressive stress, the planar symmetry
is broken, so that wrinkles appear to stabilize the system at relatively
low temperatures.\citep{pedro12b,lambin14} Rising the temperature
introduces thermal fluctuations in the system that can help to reduce
the effective stress suffered by the graphene layer, and thus to recover
the planar phase. This uncommon behavior is investigated here by atomistic
simulations.

\section{Simulation method}

Quantum path-integral and classical simulations of graphene, performed
with an in-house code, are presented as a function of temperature
($T$) and in-plane stress ($\tau$). Our goal is to show that, under
a \textit{constant} applied stress $\tau$, a planar free-standing
layer wrinkles by lowering the temperature. Wrinkling is a direct
consequence of a mechanical instability in the bending of the planar
layer.\citep{rodriguez15} In the thermodynamic limit, this instability
(spinodal point)\citep{maris91} corresponds to a vanishing surface
tension ($\sigma\equiv0$) of the layer. Our analysis will provide
insight into the dependence of the surface tension, $\sigma$, of
graphene with the applied stress, $\tau$, and the temperature.

The implementation of path-integral (PI) molecular dynamic (MD) simulations
is based on an isomorphism between the quantum system and a fictitious
classical one, in which the quantum particle (here a C nucleus) is
described by a ring polymer composed of $N_{Tr}$ (Trotter number)
beads.\citep{feynman72,ceperley95,tu02,herrero14,cazorla17} This
becomes exact in the limit $N_{Tr}\rightarrow\infty$. $N_{Tr}$ was
taken here proportional to the inverse temperature, $N_{Tr}T=6000$
K, a condition that makes the numerical error of solving the path
integral nearly temperature independent. The classical limit is achieved
just by setting $N_{Tr}=1$. The empirical interatomic LCBOPII model
was employed for the calculation of interatomic forces and potential
energy.\citep{los05} This empirical potential has been used in the
past to study the elastic behavior and the out-of-plane crumpling
of graphene.\citep{fasolino07,los16a,Los09} The phonon dispersion
curves of graphene and graphite in the harmonic limit, as derived
from the diagonalization of the dynamical matrix with the LCBOPII
model were presented in Fig. 1 of Ref. \onlinecite{karssemeijer_2011}.
The phonon dispersion of graphite displays a reasonable overall agreement
with experimental data, considering that the potential was not specifically
fitted to reproduce the force constants of graphite.\citep{karssemeijer_2011}
The largest disagreement was found for the bending rigidity $\kappa$,
that amounts to 0.7 eV in the harmonic zero-temperature limit, while
the best fit to available theoretical and experimental data reported
by Lambin amounts to 1.6 eV.\citep{lambin14} According to previous
simulations,\citep{ramirez16,ramirez17,herrero16,herrero17,herrero18}
the original LCBOPII parameterization has been slightly modified to
increase the zero-temperature bending constant of graphene to 1.5
eV.\citep{los16} For the improved potential, the phonon dispersion
curves of graphene remain unchanged for the in-plane 
modes,\citep{karssemeijer_2011}
while the improved optical (ZO) and acoustic (ZA) out-of-plane modes
were presented in Fig. 1b of Ref. \onlinecite{ramirez16}. We have
also compared the improved ZA and ZO harmonic dispersion curves of
the LCBOPII model with those corresponding to a density-functional-based
tight-binding (TB) model,\citep{porezag_95} that has been previously
employed to study the out-of-plane wrinkling of graphene.\citep{tapaszto12}
We find that the long wave-length limit of the ZA branch is nearly
identical for both methods. 

The simulations have been performed in the $N\tau T$ ensemble. The
simulation cell contains $N$ carbon atoms and 2D periodic boundary
conditions were applied with translation vectors defining the $(x,y)$-plane.
The area of the 2D simulation cell is $NA_{p}$. The in-plane stress
$\tau$ is the lateral force per unit length at the boundary of the
simulation cell.\citep{fournier08} It is defined as one-half of the
trace of the in-plane stress tensor $(\tau_{xx}+\tau_{yy})/2$. The
estimator employed for $\tau_{xx}$ and $\tau_{yy}$ can be found
in Refs. \onlinecite{ramirez16} and \onlinecite{herrero17} for classical
and quantum cases, respectively. Cells sizes with 960 atoms were studies,
and typical $N\tau T$ simulations consisted on $10^{6}$ MD steps
for equilibration and $8\times10^{6}$ steps for the calculation of
ensemble average quantities. The analysis of the simulation trajectories
was performed in subsets of 16000 different configurations, stored
at equidistant times along the simulation run. Error bars were derived
by averaging results obtained from at least two independent full trajectories.
The time step of the simulations was 1 fs. Further technical details
are identical to those already published in our previous studies of
graphene.\citep{ramirez16,ramirez17,herrero16,herrero17,herrero18}

\begin{figure}
\includegraphics[width=8.0cm]{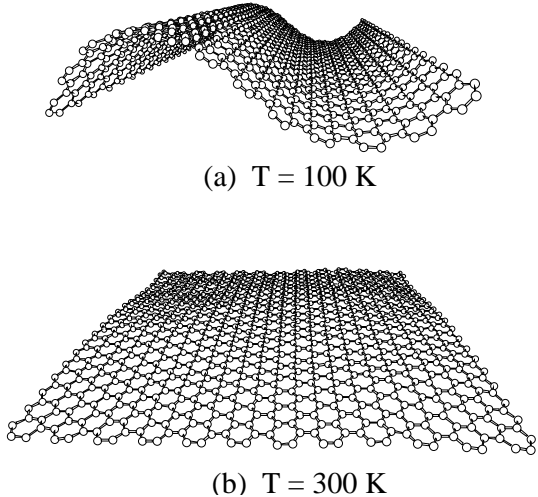}
\caption{Morphology of free-standing graphene derived from PIMD $N\tau T$
simulations at the uniform compressive in-plane stress is $\tau=0.025$
eV/$\textrm{\AA}^{2}$. The equilibrium configuration is wrinkled
at 100 K (a) but planar at 300 K (b). Wrinkles are more easily formed
along the armchair direction.}
\label{fig:morpho}
\end{figure}

\section{Layer morphology}

An example of the different morphologies found for graphene is given
in Fig. 1. These configurations were obtained by path-integral simulations
under isotropic compressive in-plane stress 
($\tau=0.025$ eV/$\textrm{\AA}^{2}$),
but at different temperatures of 100 and 300 K, respectively. The
wrinkled and planar morphologies of graphene display different values
of the projected in-plane area $NA_{p}$, while the real surface areas
$NA$ are similar (see Fig. \ref{fig:morpho}). The real surface area
per atom, $A$, is larger than $A_{p}$, if the layer is not strictly
flat. This area, $NA$, was calculated by triangulation, with six
contiguous triangles filling each hexagon of the layer. They share
a common vertex located at the barycenter of the hexagon and each
triangle has a CC bond as a side. An ongoing discussion in biological
membranes is that their thermodynamic properties should be described
using the notion of a real surface $A$ rather than an in-plane projection
$A_{p}.$\citep{chacon15} The contrast between the extensive variables,
$NA_{p}$ and $NA$,\citep{pozzo11,nicholl17} can be translated to
their conjugate intensive ones, namely the in-plane stress, $\tau$,
and the negative of the surface tension, $-\sigma,$ 
respectively.\citep{tarazona_13,fournier08,safran}
In the following, we will show that the increase of the surface tension
$\sigma$ with raising temperature drives the change from the wrinkled
to the planar morphology. This effect is a counterintuitive property
of the solid membrane. Liquid surfaces, say water, behave in the opposite
way, their surface tension $\sigma$ decreases as temperature 
increases.\citep{kayser76}

\section{Surface tension}

The calculation of the surface tension, $\sigma$, has been performed
by two routes. The first one is based on the Fourier analysis of the
amplitude of the out-of-plane atomic fluctuations in the planar morphology, 

\begin{equation}
H(\mathbf{k})=\frac{1}{N}\sum_{j=1}^{N}h_{j}e^{-i\mathbf{k}\mathbf{u}_{j}}\;.
\end{equation}
$\mathbf{k}$ is a 2D reciprocal vector commensurate with the employed
simulation cell. In the case of a classical MD simulation 
$\mathbf{r}_{j}=(\mathbf{u}_{j},h_{j})$
are the atomic positions, with $\mathbf{u}_{j}$ a 2D vector in the
$(x,y)$-plane and $h_{j}$ the height of the atom with respect to
the mean layer plane. In the case of a quantum simulation $\mathbf{r}_{j}$
are centroid coordinates, which represent the center-of-mass of the
cyclic paths associated to a given nucleus.\citep{ramirez02} The
estimation of $H(\mathbf{k})$ using centroid coordinates, instead
of bead coordinates, is justified because the centroid density represents
the static response of the quantum system to the application of an
external force.\citep{ramirez99} Thus, the mean-square amplitude
$\bar{H}^{2}=HH^{*}$ can be related to the dispersion relation, 
$\rho\omega^{2}$, of the acoustic ZA modes  as
\begin{equation}
 \left\langle \bar{H}(\mathbf{k})^{2}\right\rangle =
  \frac{k_{B}T}{A_{p}\rho[\omega(\mathbf{k})]^{2}} \; ,
\label{eq:ZA amplitude}
\end{equation}
where the angle brackets indicate an average over the whole trajectory,
$k_{B}$ is the Boltzmann constant and $\rho$ is the density of the
layer. This relation between spatial centroid fluctuations, 
$\left\langle \bar{H}^{2}\right\rangle$, and vibrational 
wavenumbers, $\omega$, has been applied in PI simulations
to study anharmonic shifts in the vibrational frequencies of solids
and molecules.\citep{ramirez01,ramirez05,herrero10} . The long-wavelength
limit of the ZA modes is well described by the dispersion relation 
\begin{equation}
  \rho[\omega(k)]^{2}=\sigma k^{2}+\kappa k^{4},\;
\label{eq:rho_w2}
\end{equation}
where $\sigma$ is the surface tension, $\kappa$ the bending constant
of the layer, and $k=|\mathbf{k}|$. Numerical details of the fit
of the simulated amplitudes $\left\langle \bar{H}^{2}\right\rangle $
to the dispersion relation $\rho\omega^{2}$, to obtain the parameters
$\sigma$ and $\kappa$, are given in Ref. \onlinecite{ramirez16}.
In particular, Fig. 1b of this reference shows that this numerical
approach accurately reproduces, in the classical low temperature limit,
the analytical dispersion curve of the ZA modes derived by diagonalization
of the vibrational dynamical matrix of the employed potential. The
dispersion relation in Eq. (\ref{eq:rho_w2}) assumes spatial isotropy
in the 2D $\mathbf{k}$-space, being accurate for 
$k\lesssim0.5\:\textrm{\AA}^{-1}$.\citep{ramirez16} 

\begin{figure}
\vspace{0.3cm}
\includegraphics[width=7.0cm]{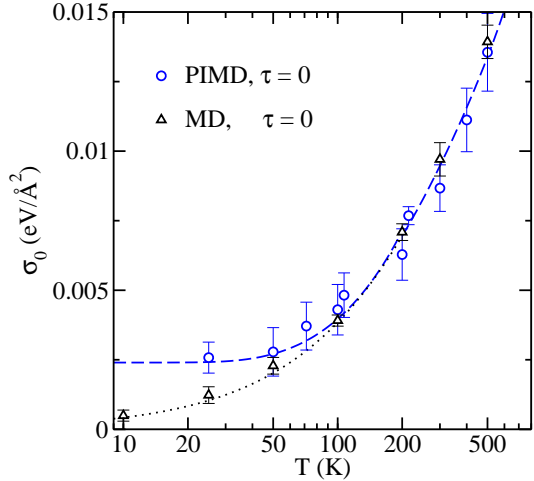}
\vspace{0.4cm}
\caption{Surface tension of free-standing
graphene as a function of temperature.
Results derived from $N\tau T$ simulations with isotropic cell fluctuations
at in-plane stress $\tau=0.$ Quantum results are shown as open circles.
Classical results are displayed by open triangles. Lines are guides
to the eye.}
\label{fig:sigma_cero}
\end{figure}

The surface tension, $\sigma_{0}\equiv\sigma(\tau=0)$, in the quantum
and classical cases has been derived from $N\tau T$ simulations with
isotropic cell fluctuations and vanishing in-plane stress $(\tau=0)$
as a function of temperature. The results for $N=960$ atoms are presented
in Fig. 2. The classical limit of $\sigma_{0}$ is somewhat larger
than that presented in Ref. \onlinecite{ramirez16}. The latter was
derived with full (i.e., non isotropic) cell fluctuations,\citep{martyna94}
allowing for an additional relaxation of the surface tension. Finite
size errors in $\sigma_{0}$ are small for the employed simulation
cell.\citep{ramirez16} The surface tension vanishes in the classical
$T\rightarrow0$ limit as the absence of bending implies that $A\equiv A_{p}$
and $-\sigma\equiv\tau$. According to Eq. (\ref{eq:rho_w2}) the
value $\sigma=0$ represents the limit for the mechanical stability
of a flat layer. For $\sigma<0$, the long-wavelength ZA modes $(k\rightarrow0)$
become mechanically unstable, as there appear imaginary wavenumbers
in $\omega(k)$.\citep{ma12,pedro12b} The classical surface tension
$\sigma$ increases with temperature, implying that the planar morphology
is stabilized,\citep{meyer06} as the dispersion relation moves away
from the mechanical instability at $\sigma=0.$

Quantum effects in $\sigma_{0}$ are significant at temperatures below
100 K. Zero-point vibrations imply a small but finite bending of the
layer in the $T\rightarrow0$ limit. The extrapolation indicates a
finite surface tension of $\sigma_{0}\sim2.5$ meV/$\textrm{\AA}^{2}$
as $T\rightarrow0$. This positive value of $\sigma_{0}$ implies
that quantum zero-point vibrations stabilize the planar morphology
of the layer with respect to the classical limit.

Our non-perturbational results for $\sigma_{0}$ are in good agreement
with recent analytical work based on a perturbational treatment of
anharmonicity in a continuous model of the solid 
membrane,\citep{adamyan16,bondarev18,amorim14,michel15}
although this term is absent in other perturbation theory 
treatment.\citep{burmistrov16}
A surface tension $\sigma$ implies a finite acoustic sound velocity
$v=(\sigma/\rho)^{1/2}$ for the out-of-plane modes. The surface tension
at 300 K is $\sigma_{0}=8.7\pm0.8$ meV/$\textrm{\AA}^{2}$, that
translates into an acoustic sound velocity of $0.4$ km/s.

In contrast to our MD results for $\sigma_{0}$, previous Monte Carlo
(MC) works claim that such term should not be present.\citep{fasolino07,hasik18}
Both MD and MC methods should provide identical results. The origin
of this disagreement is not clear, but it might be due to inaccuracies
in the sampling of the sluggish long-wavelength modes. Several general
considerations on the appearance of a finite surface tension in an
unstressed layer are appropriate. The difference between the real
and projected areas of a flat layer, $A$ and $A_{p},$ has been demonstrated
experimentally by their different stress-strain curves, \citep{nicholl17}
elastic constants, \citep{nicholl15,ramirez17} and thermal expansion
coefficients.\citep{pozzo11} It seems physically reasonable that
when the extensive variables ($NA_{p},NA$) are different, their conjugate
intensive ones, ($\tau,-\sigma$), might be also different. Time reversal
symmetry in the dynamical vibrational matrix implies that the ZA phonon
dispersion satisfies $\omega^{2}(\mathbf{k})=\omega^{2}(-\mathbf{k})$.
Eq. (\ref{eq:rho_w2}) represents the first terms of the Maclaurin
series of an analytical even phonon dispersion with coefficients depending
on ($\tau,T$). The simulations in Refs. \onlinecite{fasolino07,hasik18}
were fitted to a model following a non-analytical dispersion relation,
$\omega^{2}(\mathbf{k})\propto k^{4-\eta}$, with the anomalous exponent
$\eta\sim0.82.$ The theoretical basis of this model is the self-consistent
screening approximation (SCSA) applied to an unstressed membrane.
This model also predicts a negative Poisson ratio, $\nu$.\citep{doussal17}
However experimental data\citep{politano12} and 
computer simulations\citep{los16a}
show that $\nu\sim0.16$ for graphene. A finite term $\sigma_{0}$
is believed to be prohibited for a continuous unstressed membrane,
since it violates the rotation invariance.\citep{jiang_15} However,
in a more realistic atomistic description of a membrane a finite $\sigma_{0}$
is possible without loss of rotational invariance.\citep{kumar10} 

\begin{figure}
\vspace{0.3cm}
\includegraphics[width=7.5cm]{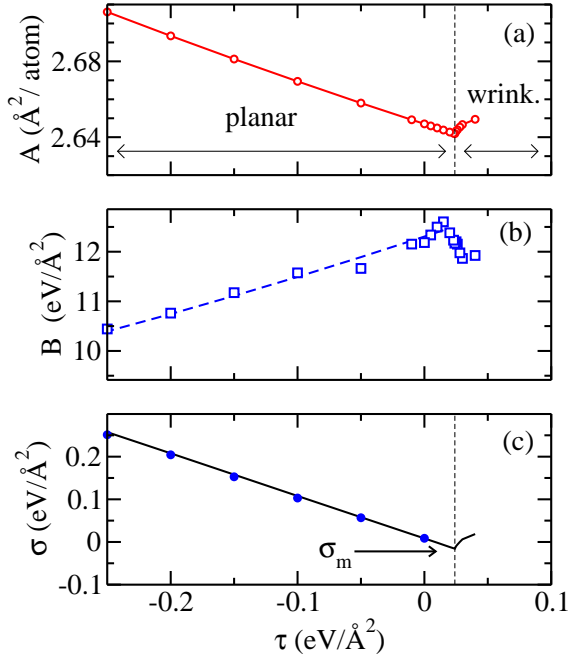}
\caption{\textit{(a)} Dependence of the real area $A$ of graphene with the
in-plane stress $\tau$ as derived from PIMD simulations at 300 K.
The solid line is a guide to the eye. The vertical dashed line indicates
the transition from a planar to a wrinkled morphology of the layer.
\textit{(b) }The corresponding 2D compressional modulus $B$ as derived
from the fluctuations of the real area $A$. The broken line is a
guide to the eye\textit{. (c) }The full line displays the surface
tension $\sigma$ obtained from the numerical integration of $B/A$\textit{
}according to Eq. (\ref{eq: sigma_int}). The full circles are obtained
from the analysis of the ZA amplitudes by Eqs. (\ref{eq:ZA amplitude})
and (\ref{eq:rho_w2}). The surface tension $\sigma$ is
minimum ($\sigma_{m}=-24$ meV / \AA$^2$)
when the layer changes its morphology.}
\label{fig:sigma_int}
\end{figure}

The consideration that $\sigma$ is the thermodynamic variable conjugate
to the real area, $NA,$ suggests a second route to calculate $\sigma$
from computer simulations.\citep{waheed09,chacon15} The average value
of the area $A$ as a function of $\tau$ is presented 
in Fig. \ref{fig:sigma_int}a
at 300 K. The real area $A$ in the planar morphology decreases when
the in-plane stress increases from tensile ($\tau<0$) to compressive
($\tau>0$) ones. We have checked that the stress-strain curve derived
from Fig. \ref{fig:sigma_cero}a is in good agreement to those derived
from Raman spectroscopy in Ref. \onlinecite{nicholl17}.\citep{ramirez18}
The planar morphology becomes unstable for the in-plane stress displayed
by a vertical broken line in Fig. \ref{fig:sigma_int}a. The change
in morphology affects the area $A$, and one observes that the slope
of $A(\tau)$ changes its sign when the layer wrinkles. The 2D modulus
of hydrostatic compression, $B$,\citep{behrrozi_96} is the inverse
of the compressibility of the real area $A$. It has been derived
from the fluctuation formula\citep{ramirez17}
\begin{equation}
 B = \frac{k_{B}T\left\langle A\right\rangle }
     {N\left(\left\langle A^{2}\right\rangle -
      \left\langle A\right\rangle \right)^{2}} \: ,
\end{equation}
and is displayed as a function of $\tau$ in Fig. \ref{fig:sigma_int}b.
The results for $A(\tau)$ and $B(\tau)$ can be combined to obtain
the surface tension $\sigma$ by numerical integration of the formal
relation between the compressional  modulus and the Hooke's law of
elasticity:\citep{waheed09,chacon15}
\begin{equation}
\frac{d\sigma}{dA}=\frac{B}{A}\:.\label{eq: sigma_int}
\end{equation}
As integration constant we used the value 
$\sigma_{0}=8.7$ meV/$\textrm{\AA}^{2}$
at $\tau=0$ (see Fig. \ref{fig:sigma_cero}). By combining the integrated
function $\sigma(A)$ and $A(\tau)$, one gets the function $\sigma(\tau)$
in Fig. \ref{fig:sigma_int}c (solid line). The surface tension attains
its minimum value $(\sigma_{m})$ when the layer becomes wrinkled.
Note the similar behavior of the conjugate variables, $\sigma$ and
$A$, with the in-plane stress $\tau$ in Fig. \ref{fig:sigma_int}a
and \ref{fig:sigma_int}c. An independent derivation of $\sigma$
from the Fourier analysis of the ZA fluctuations is shown as full
symbols for several values of $\tau$ in Fig. \ref{fig:sigma_int}c.
The agreement between both methods is excellent. For the planar morphology
at a constant temperature, the surface tension and the in-plane stress
are related as $\sigma=\sigma_{0}-\tau$.\citep{ramirez17} This relation
is not valid for the wrinkled morphology as the slope of the function
$\sigma(\tau)$ becomes positive. Compressive stresses slightly larger
than those in Fig. \ref{fig:sigma_int} produce a collapse of the
graphene structure. 

\begin{figure}
\vspace{0.2cm}
\includegraphics[width=7.0cm]{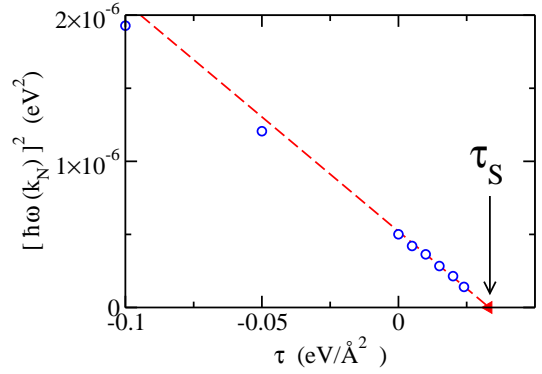}
\vspace{0.4cm}
\caption{Square of the energy quantum of the
ZA mode with lowest energy $\omega(k_{N})$
from PIMD simulations with $N=960$ atoms. The open circles were derived
at 300 K as a function of the in-plane stress $\tau$. The broken
line is a linear fit to the simulation data with
$\tau\geq10$ meV/$\textrm{\AA}^{2}$.
$\omega(k_{N})$ vanishes at the in-plane spinodal stress, $\tau_{S}=33$
meV/$\textrm{\AA}^{2}$ (full triangle). }
\label{fig:w01}
\end{figure}

The wrinkled morphology of the layer is a consequence of an instability
of the long-wavelength ZA modes,\citep{rodriguez15} whose dispersion
relation is given by Eq. (\ref{eq:rho_w2}). The energy quantum of
the ZA mode with lowest energy, $\omega(k_{N})$, of a planar layer
is plotted in Fig. \ref{fig:w01} as a function of the in-plane stress.
$k_{N}=$ $2\pi/(NA_{p})^{1/2}$ is the modulus of the $\mathbf{k}-$vector
closest to the origin. A mechanical instability appears when 
$\omega(k_{N})\rightarrow0$.\citep{ma12,pedro12b}
With the employed simulation cell, this condition is met at the spinodal
in-plane stress $\tau_{S}=33$ meV/$\textrm{\AA}^{2}$, as derived
from the extrapolation in Fig. \ref{fig:w01}. 
According to Eqs. (\ref{eq:rho_w2}),
the spinodal in-plane stress $(\tau_{S})$ and the spinodal surface
tension $(\sigma_{S})$ are related as
\begin{equation}
\sigma_{S}\equiv\sigma_{0}-\tau_{S}=-\kappa k_{N}^{2}\:.\label{eq:spinodal}
\end{equation}
Note that the r.h.s of this equation is a finite size contribution.
In the thermodynamic limit, $N\rightarrow\infty,$ then $k_{N}\rightarrow0$,
and the spinodal in-plane stress becomes $\tau_{S}=\sigma_{0}$. For
finite size systems the planar morphology is comparatively more stable,
as if $k_{N}>0$ then the spinodal surface tension will be $\sigma_{S}<0$.
For the employed simulation cell $(N=960),$ the values of $\kappa$
and $k_{N}$ derived from the PIMD trajectory at 300 K and $\tau=0$
are $\kappa=1.6$ eV, and $k_{N}$=0.123 $\textrm{\AA}^{-1}$, respectively.
Considering the value of $\tau_{S}$ from Fig. \ref{fig:w01}, one
derives from Eq. (\ref{eq:spinodal}) that 
$\sigma_{0} = 9$ meV/$\textrm{\AA}^{2}$.
This new estimation of $\sigma_{0}$ at 300 K agrees closely with
the value shown in Fig. \ref{fig:sigma_cero}.

\begin{figure}
\vspace{0.2cm}
\includegraphics[width=7.0cm]{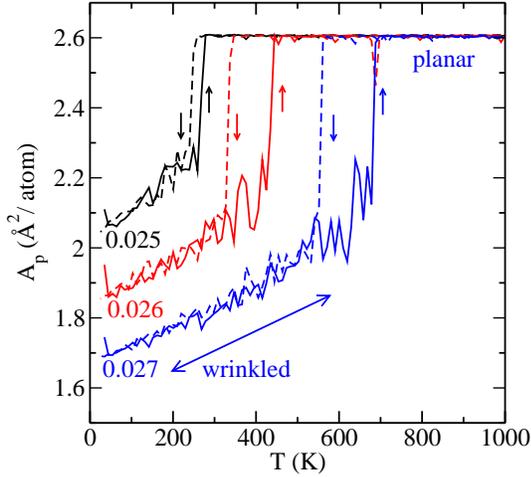}
\vspace{0.4cm}
\caption{Dependence of the projected area of graphene with temperature in a
thermal cycle between 1000 and 25 K from classical $N\tau T$ simulations
. Results for three different isotropic in-plane compressive stresses
($\tau$, in eV/$\textrm{\AA}^{2}$). Small arrows up (down) indicate
the heating (cooling) process of the cycle. }
\label{fig:ap_t}
\end{figure}

\section{Temperature cycle}

Our simulations can be compared to recent experimental data that demonstrate
the thermal control of the graphene 
morphology.\citealp{bao09,hattab11,bao12,zhang_13,bai14,deng15,meng13}
To this aim we have performed non-equilibrium simulations at constant
in-plane stress $\tau$ with temperature varying at a uniform rate
of 20 K/ns in cycles between 1000 and 25 K. A cycle consists of $10^{8}$
MD steps. The simulations are performed in the classical limit, as
quantum effects in the morphology of the layer are significant only
below 100 K. The morphology has been monitored by the value of the
projected area, $A_{p}$, along the thermal cycle. The results for
three different in-plane compressive stresses are displayed in Fig.
\ref{fig:ap_t}. An arrow pointing down (up) indicates that the temperature
is decreasing (increasing). At high temperature the in-plane area
has a value of $A_{p}\sim2.6$ $\textrm{\AA}^{2}$/atom, typical of
a planar morphology. At the scale of the figure, the area $A_{p}$
remains nearly constant as the temperature decreases. We observe that
by cooling the flat layer becomes wrinkled and the projected area
$A_{p}$ jumps to a value $<2$ $\textrm{\AA}^{2}$/atom. The lower
the compressive in-plane stress, $\tau$, the lower the temperature
of the wrinkling transition. By cooling down to 25 K the projected
area of the wrinkled morphology decreases monotonically, showing that
the lower the temperature the larger the amplitude of the surface
wrinkles of a free-standing layer.\citep{kirilenko11} The area $A_{p}$
in the wrinkled morphology is extremely sensitive to the applied in-plane
stress. Therefore, the strain in the variable $A_{p}$ for the wrinkled
morphology should be strongly dependent on external conditions, such
as the substrate and the size of the sample. A comparison to experimental
data for this variable is only sensible in a qualitative level. From
an experimental point of view, there appears an ample range of periodic
wrinkling morphologies, from amplitudes of 1 $\textrm{\AA\ and 
wavelengths of 8 }$\AA ,\citep{tapaszto12}
to amplitudes of 300 \AA{} and wavelengths of 25 $\mu\textrm{m}$.\citep{bao09}
In the reverse cycle, upon heating from 25 K up to 1000 K, one observes
that the change in morphology is reversible and there appears hysteresis
in the transition temperature.

\section{Summary}

Summarizing, the surface tension $\sigma$ of graphene, as the variable
conjugate to the real surface area, $A$, has been determined by three
different methods: by analysis of the Fourier transform of the bending
fluctuations, by integration of the 2D compressional modulus associated
to the real area $A,$ and by derivation of the spinodal in-plane
stress $(\tau_{S})$ in a finite size sample. The mutual agreement
reveals that our analysis is thermodynamically sound, providing new
insight into this intrinsic property. The consideration that the bending
of a planar layer increases its surface tension allows us to rationalize
that quantum zero-point effects as well as a rise of temperature increase
the stability of a planar morphology. Quantum effects in the surface
tension are significant below 100 K. 

Our simulations provide new insight into experiments showing a thermal
control of the graphene 
morphology.\citealp{bao09,hattab11,bao12,zhang_13,bai14,deng15,meng13}
The temperature changes the bending of the layer while the latter
modifies the surface tension. The higher the temperature, the larger
the surface tension, favoring a planar layer. The decrease of the
surface tension with lowering temperature produces wrinkles when the
planar layer approaches its stability limit (spinodal point). This
transition is reversible and shows hysteresis in agreement to experiments
performed on supported 
graphene.\citealp{bao09,hattab11,bao12,zhang_13,bai14,deng15,meng13}
The mechanical instability in the bending of the planar layer displays
a size effect. The cut-off of the long-wavelength bending modes in
a finite size layer implies an increased stability of its planar morphology.
In the thermodynamic limit $N\rightarrow\infty$, the spinodal point
of a planar layer corresponds to a vanishing surface tension. The
expectation that a membrane in thermal equilibrium has vanishing surface
tension, because its free energy should be minimal with respect to
the area of the membrane, is not met for crystalline graphene. This
is a consequence of the coupling between the real surface area and
the bending of the layer.

\acknowledgments 

This work was supported by Dirección General de Investigación, MINECO
(Spain) through Grant No. FIS2015-64222-C2-1-P. We thank the support
of J. H. Los in the implementation of the LCBOPII potential.


\begin{thebibliography}{61}
\expandafter\ifx\csname natexlab\endcsname\relax\def\natexlab#1{#1}\fi
\expandafter\ifx\csname bibnamefont\endcsname\relax
  \def\bibnamefont#1{#1}\fi
\expandafter\ifx\csname bibfnamefont\endcsname\relax
  \def\bibfnamefont#1{#1}\fi
\expandafter\ifx\csname citenamefont\endcsname\relax
  \def\citenamefont#1{#1}\fi
\expandafter\ifx\csname url\endcsname\relax
  \def\url#1{\texttt{#1}}\fi
\expandafter\ifx\csname urlprefix\endcsname\relax\def\urlprefix{URL }\fi
\providecommand{\bibinfo}[2]{#2}
\providecommand{\eprint}[2][]{\url{#2}}

\bibitem[{\citenamefont{Roldan et~al.}(2017)\citenamefont{Roldan, Chirolli,
  Prada, Angel Silva-Guillen, San-Jose, and Guinea}}]{roldan17}
\bibinfo{author}{\bibfnamefont{R.}~\bibnamefont{Roldan}},
  \bibinfo{author}{\bibfnamefont{L.}~\bibnamefont{Chirolli}},
  \bibinfo{author}{\bibfnamefont{E.}~\bibnamefont{Prada}},
  \bibinfo{author}{\bibfnamefont{J.}~\bibnamefont{Angel Silva-Guillen}},
  \bibinfo{author}{\bibfnamefont{P.}~\bibnamefont{San-Jose}}, \bibnamefont{and}
  \bibinfo{author}{\bibfnamefont{F.}~\bibnamefont{Guinea}},
  \bibinfo{journal}{Chem. Soc. Rev.} \textbf{\bibinfo{volume}{46}},
  \bibinfo{pages}{4387} (\bibinfo{year}{2017}).

\bibitem[{\citenamefont{Amorim et~al.}(2016)\citenamefont{Amorim, Cortijo,
  de~Juan, Grushin, Guinea, Guti\'errez-Rubio, Ochoa, Parente, Rold\'an,
  San-Jose et~al.}}]{amorim16}
\bibinfo{author}{\bibfnamefont{B.}~\bibnamefont{Amorim}},
  \bibinfo{author}{\bibfnamefont{A.}~\bibnamefont{Cortijo}},
  \bibinfo{author}{\bibfnamefont{F.}~\bibnamefont{de~Juan}},
  \bibinfo{author}{\bibfnamefont{A.}~\bibnamefont{Grushin}},
  \bibinfo{author}{\bibfnamefont{F.}~\bibnamefont{Guinea}},
  \bibinfo{author}{\bibfnamefont{A.}~\bibnamefont{Guti\'errez-Rubio}},
  \bibinfo{author}{\bibfnamefont{H.}~\bibnamefont{Ochoa}},
  \bibinfo{author}{\bibfnamefont{V.}~\bibnamefont{Parente}},
  \bibinfo{author}{\bibfnamefont{R.}~\bibnamefont{Rold\'an}},
  \bibinfo{author}{\bibfnamefont{P.}~\bibnamefont{San-Jose}},
  \bibnamefont{et~al.}, \bibinfo{journal}{Phys. Reports}
  \textbf{\bibinfo{volume}{617}}, \bibinfo{pages}{1 } (\bibinfo{year}{2016}).

\bibitem[{\citenamefont{Bao et~al.}(2009)\citenamefont{Bao, Miao, Chen, Zhang,
  Jang, Dames, and Lau}}]{bao09}
\bibinfo{author}{\bibfnamefont{W.}~\bibnamefont{Bao}},
  \bibinfo{author}{\bibfnamefont{F.}~\bibnamefont{Miao}},
  \bibinfo{author}{\bibfnamefont{Z.}~\bibnamefont{Chen}},
  \bibinfo{author}{\bibfnamefont{H.}~\bibnamefont{Zhang}},
  \bibinfo{author}{\bibfnamefont{W.}~\bibnamefont{Jang}},
  \bibinfo{author}{\bibfnamefont{C.}~\bibnamefont{Dames}}, \bibnamefont{and}
  \bibinfo{author}{\bibfnamefont{C.~N.} \bibnamefont{Lau}},
  \bibinfo{journal}{Nature Nanotechnol.} \textbf{\bibinfo{volume}{9}},
  \bibinfo{pages}{562} (\bibinfo{year}{2009}).

\bibitem[{\citenamefont{Hattab et~al.}(2012)\citenamefont{Hattab, N'Diaye,
  Wall, Klein, Jnawali, Coraux, Busse, van Gastel, Poelsema, Michely
  et~al.}}]{hattab11}
\bibinfo{author}{\bibfnamefont{H.}~\bibnamefont{Hattab}},
  \bibinfo{author}{\bibfnamefont{A.~T.} \bibnamefont{N'Diaye}},
  \bibinfo{author}{\bibfnamefont{D.}~\bibnamefont{Wall}},
  \bibinfo{author}{\bibfnamefont{C.}~\bibnamefont{Klein}},
  \bibinfo{author}{\bibfnamefont{G.}~\bibnamefont{Jnawali}},
  \bibinfo{author}{\bibfnamefont{J.}~\bibnamefont{Coraux}},
  \bibinfo{author}{\bibfnamefont{C.}~\bibnamefont{Busse}},
  \bibinfo{author}{\bibfnamefont{R.}~\bibnamefont{van Gastel}},
  \bibinfo{author}{\bibfnamefont{B.}~\bibnamefont{Poelsema}},
  \bibinfo{author}{\bibfnamefont{T.}~\bibnamefont{Michely}},
  \bibnamefont{et~al.}, \bibinfo{journal}{Nano Letters}
  \textbf{\bibinfo{volume}{12}}, \bibinfo{pages}{678} (\bibinfo{year}{2012}).

\bibitem[{\citenamefont{Bao et~al.}(2012)\citenamefont{Bao, Myhro, Zhao, Chen,
  Jang, Jing, Miao, Zhang, Dames, and Lau}}]{bao12}
\bibinfo{author}{\bibfnamefont{W.}~\bibnamefont{Bao}},
  \bibinfo{author}{\bibfnamefont{K.}~\bibnamefont{Myhro}},
  \bibinfo{author}{\bibfnamefont{Z.}~\bibnamefont{Zhao}},
  \bibinfo{author}{\bibfnamefont{Z.}~\bibnamefont{Chen}},
  \bibinfo{author}{\bibfnamefont{W.}~\bibnamefont{Jang}},
  \bibinfo{author}{\bibfnamefont{L.}~\bibnamefont{Jing}},
  \bibinfo{author}{\bibfnamefont{F.}~\bibnamefont{Miao}},
  \bibinfo{author}{\bibfnamefont{H.}~\bibnamefont{Zhang}},
  \bibinfo{author}{\bibfnamefont{C.}~\bibnamefont{Dames}}, \bibnamefont{and}
  \bibinfo{author}{\bibfnamefont{C.~N.} \bibnamefont{Lau}},
  \bibinfo{journal}{Nano Letters} \textbf{\bibinfo{volume}{12}},
  \bibinfo{pages}{5470} (\bibinfo{year}{2012}).

\bibitem[{\citenamefont{Zhang et~al.}(2013)\citenamefont{Zhang, Fu, Cui, Mu,
  Jin, and Bao}}]{zhang_13}
\bibinfo{author}{\bibfnamefont{Y.}~\bibnamefont{Zhang}},
  \bibinfo{author}{\bibfnamefont{Q.}~\bibnamefont{Fu}},
  \bibinfo{author}{\bibfnamefont{Y.}~\bibnamefont{Cui}},
  \bibinfo{author}{\bibfnamefont{R.}~\bibnamefont{Mu}},
  \bibinfo{author}{\bibfnamefont{L.}~\bibnamefont{Jin}}, \bibnamefont{and}
  \bibinfo{author}{\bibfnamefont{X.}~\bibnamefont{Bao}},
  \bibinfo{journal}{Phys. Chem. Chem. Phys.} \textbf{\bibinfo{volume}{15}},
  \bibinfo{pages}{19042} (\bibinfo{year}{2013}).

\bibitem[{\citenamefont{Bai et~al.}(2014)\citenamefont{Bai, Zhou, Zheng, Meng,
  Peng, Liu, Nie, and He}}]{bai14}
\bibinfo{author}{\bibfnamefont{K.-K.} \bibnamefont{Bai}},
  \bibinfo{author}{\bibfnamefont{Y.}~\bibnamefont{Zhou}},
  \bibinfo{author}{\bibfnamefont{H.}~\bibnamefont{Zheng}},
  \bibinfo{author}{\bibfnamefont{L.}~\bibnamefont{Meng}},
  \bibinfo{author}{\bibfnamefont{H.}~\bibnamefont{Peng}},
  \bibinfo{author}{\bibfnamefont{Z.}~\bibnamefont{Liu}},
  \bibinfo{author}{\bibfnamefont{J.-C.} \bibnamefont{Nie}}, \bibnamefont{and}
  \bibinfo{author}{\bibfnamefont{L.}~\bibnamefont{He}}, \bibinfo{journal}{Phys.
  Rev. Lett.} \textbf{\bibinfo{volume}{113}}, \bibinfo{pages}{086102}
  (\bibinfo{year}{2014}),
  \urlprefix\url{https://link.aps.org/doi/10.1103/PhysRevLett.113.086102}.

\bibitem[{\citenamefont{Deng and Berry}(2016)}]{deng15}
\bibinfo{author}{\bibfnamefont{S.}~\bibnamefont{Deng}} \bibnamefont{and}
  \bibinfo{author}{\bibfnamefont{V.}~\bibnamefont{Berry}},
  \bibinfo{journal}{Materials Today} \textbf{\bibinfo{volume}{19}},
  \bibinfo{pages}{197 } (\bibinfo{year}{2016}), ISSN \bibinfo{issn}{1369-7021},
  \urlprefix\url{http://www.sciencedirect.com/science/article/pii/S1369702115003119}.

\bibitem[{\citenamefont{Meng et~al.}(2013)\citenamefont{Meng, Su, Geng, Yu,
  Liu, Dou, Nie, and He}}]{meng13}
\bibinfo{author}{\bibfnamefont{L.}~\bibnamefont{Meng}},
  \bibinfo{author}{\bibfnamefont{Y.}~\bibnamefont{Su}},
  \bibinfo{author}{\bibfnamefont{D.}~\bibnamefont{Geng}},
  \bibinfo{author}{\bibfnamefont{G.}~\bibnamefont{Yu}},
  \bibinfo{author}{\bibfnamefont{Y.}~\bibnamefont{Liu}},
  \bibinfo{author}{\bibfnamefont{R.-F.} \bibnamefont{Dou}},
  \bibinfo{author}{\bibfnamefont{J.-C.} \bibnamefont{Nie}}, \bibnamefont{and}
  \bibinfo{author}{\bibfnamefont{L.}~\bibnamefont{He}}, \bibinfo{journal}{Appl.
  Phys. Lett.} \textbf{\bibinfo{volume}{103}}, \bibinfo{pages}{251610}
  (\bibinfo{year}{2013}).

\bibitem[{\citenamefont{de~Andres et~al.}(2012)\citenamefont{de~Andres, Guinea,
  and Katsnelson}}]{pedro12b}
\bibinfo{author}{\bibfnamefont{P.~L.} \bibnamefont{de~Andres}},
  \bibinfo{author}{\bibfnamefont{F.}~\bibnamefont{Guinea}}, \bibnamefont{and}
  \bibinfo{author}{\bibfnamefont{M.~I.} \bibnamefont{Katsnelson}},
  \bibinfo{journal}{Phys. Rev. B} \textbf{\bibinfo{volume}{86}},
  \bibinfo{pages}{245409} (\bibinfo{year}{2012}).

\bibitem[{\citenamefont{Lambin}(2014)}]{lambin14}
\bibinfo{author}{\bibfnamefont{P.}~\bibnamefont{Lambin}},
  \bibinfo{journal}{Appl. Sci.} \textbf{\bibinfo{volume}{4}},
  \bibinfo{pages}{282} (\bibinfo{year}{2014}).

\bibitem[{\citenamefont{Rodr\'{\i}guez-Hern\'andez}(2015)}]{rodriguez15}
\bibinfo{author}{\bibfnamefont{J.}~\bibnamefont{Rodr\'{\i}guez-Hern\'andez}},
  \bibinfo{journal}{Prog. Polym. Sci.} \textbf{\bibinfo{volume}{42}},
  \bibinfo{pages}{1 } (\bibinfo{year}{2015}), ISSN \bibinfo{issn}{0079-6700}.

\bibitem[{\citenamefont{Maris}(1991)}]{maris91}
\bibinfo{author}{\bibfnamefont{H.~J.} \bibnamefont{Maris}},
  \bibinfo{journal}{Phys. Rev. Lett.} \textbf{\bibinfo{volume}{66}},
  \bibinfo{pages}{45} (\bibinfo{year}{1991}).

\bibitem[{\citenamefont{Feynman}(1972)}]{feynman72}
\bibinfo{author}{\bibfnamefont{R.~P.} \bibnamefont{Feynman}},
  \emph{\bibinfo{title}{Statistical Mechanics}}
  (\bibinfo{publisher}{Addison-Wesley}, \bibinfo{address}{New York},
  \bibinfo{year}{1972}).

\bibitem[{\citenamefont{Ceperley}(1995)}]{ceperley95}
\bibinfo{author}{\bibfnamefont{D.~M.} \bibnamefont{Ceperley}},
  \bibinfo{journal}{Rev. Mod. Phys.} \textbf{\bibinfo{volume}{67}},
  \bibinfo{pages}{279} (\bibinfo{year}{1995}).

\bibitem[{\citenamefont{Tuckerman}(2002)}]{tu02}
\bibinfo{author}{\bibfnamefont{M.~E.} \bibnamefont{Tuckerman}}, in
  \emph{\bibinfo{booktitle}{Quantum Simulations of Complex Many--Body Systems:
  From Theory to Algorithms}}, edited by
  \bibinfo{editor}{\bibfnamefont{J.}~\bibnamefont{Grotendorst}},
  \bibinfo{editor}{\bibfnamefont{D.}~\bibnamefont{Marx}}, \bibnamefont{and}
  \bibinfo{editor}{\bibfnamefont{A.}~\bibnamefont{Muramatsu}}
  (\bibinfo{publisher}{NIC}, \bibinfo{address}{FZ J\"ulich},
  \bibinfo{year}{2002}), p. \bibinfo{pages}{269}.

\bibitem[{\citenamefont{Herrero and Ram\'irez}(2014)}]{herrero14}
\bibinfo{author}{\bibfnamefont{C.~P.} \bibnamefont{Herrero}} \bibnamefont{and}
  \bibinfo{author}{\bibfnamefont{R.}~\bibnamefont{Ram\'irez}},
  \bibinfo{journal}{J. Phys.: Condens. Matter} \textbf{\bibinfo{volume}{26}},
  \bibinfo{pages}{233201} (\bibinfo{year}{2014}).

\bibitem[{\citenamefont{Cazorla and Boronat}(2017)}]{cazorla17}
\bibinfo{author}{\bibfnamefont{C.}~\bibnamefont{Cazorla}} \bibnamefont{and}
  \bibinfo{author}{\bibfnamefont{J.}~\bibnamefont{Boronat}},
  \bibinfo{journal}{Rev. Mod. Phys.} \textbf{\bibinfo{volume}{89}},
  \bibinfo{pages}{035003} (\bibinfo{year}{2017}).

\bibitem[{\citenamefont{Los et~al.}(2005)\citenamefont{Los, Ghiringhelli,
  Meijer, and Fasolino}}]{los05}
\bibinfo{author}{\bibfnamefont{J.~H.} \bibnamefont{Los}},
  \bibinfo{author}{\bibfnamefont{L.~M.} \bibnamefont{Ghiringhelli}},
  \bibinfo{author}{\bibfnamefont{E.~J.} \bibnamefont{Meijer}},
  \bibnamefont{and} \bibinfo{author}{\bibfnamefont{A.}~\bibnamefont{Fasolino}},
  \bibinfo{journal}{Phys. Rev. B} \textbf{\bibinfo{volume}{72}},
  \bibinfo{pages}{214102} (\bibinfo{year}{2005}).

\bibitem[{\citenamefont{Fasolino et~al.}(2007)\citenamefont{Fasolino, Los, and
  Katsnelson}}]{fasolino07}
\bibinfo{author}{\bibfnamefont{A.}~\bibnamefont{Fasolino}},
  \bibinfo{author}{\bibfnamefont{J.~H.} \bibnamefont{Los}}, \bibnamefont{and}
  \bibinfo{author}{\bibfnamefont{M.~I.} \bibnamefont{Katsnelson}},
  \bibinfo{journal}{Nature Mater.} \textbf{\bibinfo{volume}{6}},
  \bibinfo{pages}{858} (\bibinfo{year}{2007}).

\bibitem[{\citenamefont{Los et~al.}(2016)\citenamefont{Los, Fasolino, and
  Katsnelson}}]{los16a}
\bibinfo{author}{\bibfnamefont{J.~H.} \bibnamefont{Los}},
  \bibinfo{author}{\bibfnamefont{A.}~\bibnamefont{Fasolino}}, \bibnamefont{and}
  \bibinfo{author}{\bibfnamefont{M.~I.} \bibnamefont{Katsnelson}},
  \bibinfo{journal}{Phys. Rev. Lett.} \textbf{\bibinfo{volume}{116}},
  \bibinfo{pages}{015901} (\bibinfo{year}{2016}).

\bibitem[{\citenamefont{Los et~al.}(2009)\citenamefont{Los, Katsnelson, Yazyev,
  Zakharchenko, and Fasolino}}]{Los09}
\bibinfo{author}{\bibfnamefont{J.~H.} \bibnamefont{Los}},
  \bibinfo{author}{\bibfnamefont{M.~I.} \bibnamefont{Katsnelson}},
  \bibinfo{author}{\bibfnamefont{O.~V.} \bibnamefont{Yazyev}},
  \bibinfo{author}{\bibfnamefont{K.~V.} \bibnamefont{Zakharchenko}},
  \bibnamefont{and} \bibinfo{author}{\bibfnamefont{A.}~\bibnamefont{Fasolino}},
  \bibinfo{journal}{Phys. Rev. B} \textbf{\bibinfo{volume}{80}},
  \bibinfo{pages}{121405} (\bibinfo{year}{2009}).

\bibitem[{\citenamefont{Karssemeijer and Fasolino}(2011)}]{karssemeijer_2011}
\bibinfo{author}{\bibfnamefont{L.}~\bibnamefont{Karssemeijer}}
  \bibnamefont{and} \bibinfo{author}{\bibfnamefont{A.}~\bibnamefont{Fasolino}},
  \bibinfo{journal}{Surface Science} \textbf{\bibinfo{volume}{605}},
  \bibinfo{pages}{1611 } (\bibinfo{year}{2011}), ISSN
  \bibinfo{issn}{0039-6028}.

\bibitem[{\citenamefont{Ram\'{\i}rez et~al.}(2016)\citenamefont{Ram\'{\i}rez,
  Chac\'on, and Herrero}}]{ramirez16}
\bibinfo{author}{\bibfnamefont{R.}~\bibnamefont{Ram\'{\i}rez}},
  \bibinfo{author}{\bibfnamefont{E.}~\bibnamefont{Chac\'on}}, \bibnamefont{and}
  \bibinfo{author}{\bibfnamefont{C.~P.} \bibnamefont{Herrero}},
  \bibinfo{journal}{Phys. Rev. B} \textbf{\bibinfo{volume}{93}},
  \bibinfo{pages}{235419} (\bibinfo{year}{2016}).

\bibitem[{\citenamefont{Ram\'{\i}rez and Herrero}(2017)}]{ramirez17}
\bibinfo{author}{\bibfnamefont{R.}~\bibnamefont{Ram\'{\i}rez}}
  \bibnamefont{and} \bibinfo{author}{\bibfnamefont{C.~P.}
  \bibnamefont{Herrero}}, \bibinfo{journal}{Phys. Rev. B}
  \textbf{\bibinfo{volume}{95}}, \bibinfo{pages}{045423}
  (\bibinfo{year}{2017}),
  \urlprefix\url{https://link.aps.org/doi/10.1103/PhysRevB.95.045423}.

\bibitem[{\citenamefont{Herrero and Ram\'{\i}rez}(2016)}]{herrero16}
\bibinfo{author}{\bibfnamefont{C.~P.} \bibnamefont{Herrero}} \bibnamefont{and}
  \bibinfo{author}{\bibfnamefont{R.}~\bibnamefont{Ram\'{\i}rez}},
  \bibinfo{journal}{J. Chem. Phys.} \textbf{\bibinfo{volume}{145}},
  \bibinfo{pages}{224701} (\bibinfo{year}{2016}).

\bibitem[{\citenamefont{Herrero and Ram\'{\i}rez}(2017)}]{herrero17}
\bibinfo{author}{\bibfnamefont{C.~P.} \bibnamefont{Herrero}} \bibnamefont{and}
  \bibinfo{author}{\bibfnamefont{R.}~\bibnamefont{Ram\'{\i}rez}},
  \bibinfo{journal}{Phys. Chem. Chem. Phys.} \textbf{\bibinfo{volume}{19}},
  \bibinfo{pages}{31898} (\bibinfo{year}{2017}),
  \urlprefix\url{http://dx.doi.org/10.1039/C7CP06821B}.

\bibitem[{\citenamefont{Herrero and Ram\'{\i}rez}(2018)}]{herrero18}
\bibinfo{author}{\bibfnamefont{C.~P.} \bibnamefont{Herrero}} \bibnamefont{and}
  \bibinfo{author}{\bibfnamefont{R.}~\bibnamefont{Ram\'{\i}rez}},
  \bibinfo{journal}{J. Chem. Phys.} \textbf{\bibinfo{volume}{148}},
  \bibinfo{pages}{102302} (\bibinfo{year}{2018}).

\bibitem[{\citenamefont{Los}(2016)}]{los16}
\bibinfo{author}{\bibfnamefont{J.~H.} \bibnamefont{Los}}
  (\bibinfo{year}{2016}), \bibinfo{note}{private communication}.

\bibitem[{\citenamefont{Porezag et~al.}(1995)\citenamefont{Porezag, Frauenheim,
  K\"ohler, Seifert, and Kaschner}}]{porezag_95}
\bibinfo{author}{\bibfnamefont{D.}~\bibnamefont{Porezag}},
  \bibinfo{author}{\bibfnamefont{T.}~\bibnamefont{Frauenheim}},
  \bibinfo{author}{\bibfnamefont{T.}~\bibnamefont{K\"ohler}},
  \bibinfo{author}{\bibfnamefont{G.}~\bibnamefont{Seifert}}, \bibnamefont{and}
  \bibinfo{author}{\bibfnamefont{R.}~\bibnamefont{Kaschner}},
  \bibinfo{journal}{Phys. Rev. B} \textbf{\bibinfo{volume}{51}},
  \bibinfo{pages}{12947} (\bibinfo{year}{1995}).

\bibitem[{\citenamefont{Tapaszt\'o et~al.}(2012)\citenamefont{Tapaszt\'o,
  Dumitric\ifmmode~\check{s}\else \v{a}\fi{}, Kim, Nemes-Incze, Hwang, and
  Bir\'o}}]{tapaszto12}
\bibinfo{author}{\bibfnamefont{L.}~\bibnamefont{Tapaszt\'o}},
  \bibinfo{author}{\bibfnamefont{T.}~\bibnamefont{Dumitric\ifmmode~\check{s}\else
  \v{a}\fi{}}}, \bibinfo{author}{\bibfnamefont{S.~J.} \bibnamefont{Kim}},
  \bibinfo{author}{\bibfnamefont{P.}~\bibnamefont{Nemes-Incze}},
  \bibinfo{author}{\bibfnamefont{C.}~\bibnamefont{Hwang}}, \bibnamefont{and}
  \bibinfo{author}{\bibfnamefont{L.~P.} \bibnamefont{Bir\'o}},
  \bibinfo{journal}{Nature Phys.} \textbf{\bibinfo{volume}{8}},
  \bibinfo{pages}{739} (\bibinfo{year}{2012}).

\bibitem[{\citenamefont{Fournier and Barbetta}(2008)}]{fournier08}
\bibinfo{author}{\bibfnamefont{J.-B.} \bibnamefont{Fournier}} \bibnamefont{and}
  \bibinfo{author}{\bibfnamefont{C.}~\bibnamefont{Barbetta}},
  \bibinfo{journal}{Phys. Rev. Lett.} \textbf{\bibinfo{volume}{100}},
  \bibinfo{pages}{078103} (\bibinfo{year}{2008}).

\bibitem[{\citenamefont{Chac\'on et~al.}(2015)\citenamefont{Chac\'on, Tarazona,
  and Bresme}}]{chacon15}
\bibinfo{author}{\bibfnamefont{E.}~\bibnamefont{Chac\'on}},
  \bibinfo{author}{\bibfnamefont{P.}~\bibnamefont{Tarazona}}, \bibnamefont{and}
  \bibinfo{author}{\bibfnamefont{F.}~\bibnamefont{Bresme}},
  \bibinfo{journal}{J. Chem. Phys.} \textbf{\bibinfo{volume}{143}},
  \bibinfo{eid}{034706} (\bibinfo{year}{2015}).

\bibitem[{\citenamefont{Pozzo et~al.}(2011)\citenamefont{Pozzo, Alf\`e,
  Lacovig, Hofmann, Lizzit, and Baraldi}}]{pozzo11}
\bibinfo{author}{\bibfnamefont{M.}~\bibnamefont{Pozzo}},
  \bibinfo{author}{\bibfnamefont{D.}~\bibnamefont{Alf\`e}},
  \bibinfo{author}{\bibfnamefont{P.}~\bibnamefont{Lacovig}},
  \bibinfo{author}{\bibfnamefont{P.}~\bibnamefont{Hofmann}},
  \bibinfo{author}{\bibfnamefont{S.}~\bibnamefont{Lizzit}}, \bibnamefont{and}
  \bibinfo{author}{\bibfnamefont{A.}~\bibnamefont{Baraldi}},
  \bibinfo{journal}{Phys. Rev. Lett.} \textbf{\bibinfo{volume}{106}},
  \bibinfo{pages}{135501} (\bibinfo{year}{2011}).

\bibitem[{\citenamefont{Nicholl et~al.}(2017)\citenamefont{Nicholl, Lavrik,
  Vlassiouk, Srijanto, and Bolotin}}]{nicholl17}
\bibinfo{author}{\bibfnamefont{R.~J.~T.} \bibnamefont{Nicholl}},
  \bibinfo{author}{\bibfnamefont{N.~V.} \bibnamefont{Lavrik}},
  \bibinfo{author}{\bibfnamefont{I.}~\bibnamefont{Vlassiouk}},
  \bibinfo{author}{\bibfnamefont{B.~R.} \bibnamefont{Srijanto}},
  \bibnamefont{and} \bibinfo{author}{\bibfnamefont{K.~I.}
  \bibnamefont{Bolotin}}, \bibinfo{journal}{Phys. Rev. Lett.}
  \textbf{\bibinfo{volume}{118}}, \bibinfo{pages}{266101}
  (\bibinfo{year}{2017}).

\bibitem[{\citenamefont{Tarazona et~al.}(2013)\citenamefont{Tarazona, Chac\'on,
  and Bresme}}]{tarazona_13}
\bibinfo{author}{\bibfnamefont{P.}~\bibnamefont{Tarazona}},
  \bibinfo{author}{\bibfnamefont{E.}~\bibnamefont{Chac\'on}}, \bibnamefont{and}
  \bibinfo{author}{\bibfnamefont{F.}~\bibnamefont{Bresme}},
  \bibinfo{journal}{J. Chem. Phys.} \textbf{\bibinfo{volume}{139}},
  \bibinfo{eid}{094902} (\bibinfo{year}{2013}).

\bibitem[{\citenamefont{Safran}(1994)}]{safran}
\bibinfo{author}{\bibfnamefont{S.~A.} \bibnamefont{Safran}},
  \emph{\bibinfo{title}{Statistical Thermodynamics of Surfaces, Interfaces, and
  Membranes}} (\bibinfo{publisher}{Addison-Wesley Reading, Massachusetts},
  \bibinfo{year}{1994}).

\bibitem[{\citenamefont{Kayser}(1976)}]{kayser76}
\bibinfo{author}{\bibfnamefont{W.~V.} \bibnamefont{Kayser}},
  \bibinfo{journal}{J. Colloid Interface Sci.} \textbf{\bibinfo{volume}{56}},
  \bibinfo{pages}{622 } (\bibinfo{year}{1976}), ISSN \bibinfo{issn}{0021-9797},
  \urlprefix\url{http://www.sciencedirect.com/science/article/pii/0021979776901302}.

\bibitem[{\citenamefont{Ram\'irez and L\'opez-Ciudad}(2002)}]{ramirez02}
\bibinfo{author}{\bibfnamefont{R.}~\bibnamefont{Ram\'irez}} \bibnamefont{and}
  \bibinfo{author}{\bibfnamefont{T.}~\bibnamefont{L\'opez-Ciudad}}, in
  \emph{\bibinfo{booktitle}{Quantum Simulations of Complex Many--Body Systems:
  From Theory to Algorithms}}, edited by
  \bibinfo{editor}{\bibfnamefont{J.}~\bibnamefont{Grotendorst}},
  \bibinfo{editor}{\bibfnamefont{D.}~\bibnamefont{Marx}}, \bibnamefont{and}
  \bibinfo{editor}{\bibfnamefont{A.}~\bibnamefont{Muramatsu}}
  (\bibinfo{publisher}{NIC}, \bibinfo{address}{FZ J\"ulich},
  \bibinfo{year}{2002}), p. \bibinfo{pages}{325}.

\bibitem[{\citenamefont{Ram\'{\i}rez and L\'opez-Ciudad}(1999)}]{ramirez99}
\bibinfo{author}{\bibfnamefont{R.}~\bibnamefont{Ram\'{\i}rez}}
  \bibnamefont{and}
  \bibinfo{author}{\bibfnamefont{T.}~\bibnamefont{L\'opez-Ciudad}},
  \bibinfo{journal}{Phys. Rev. Lett.} \textbf{\bibinfo{volume}{83}},
  \bibinfo{pages}{4456} (\bibinfo{year}{1999}).

\bibitem[{\citenamefont{Ram\'irez and L\'opez-Ciudad}(2001)}]{ramirez01}
\bibinfo{author}{\bibfnamefont{R.}~\bibnamefont{Ram\'irez}} \bibnamefont{and}
  \bibinfo{author}{\bibfnamefont{T.}~\bibnamefont{L\'opez-Ciudad}},
  \bibinfo{journal}{J. Chem. Phys.} \textbf{\bibinfo{volume}{115}},
  \bibinfo{pages}{103} (\bibinfo{year}{2001}).

\bibitem[{\citenamefont{Ram\'irez and Herrero}(2005)}]{ramirez05}
\bibinfo{author}{\bibfnamefont{R.}~\bibnamefont{Ram\'irez}} \bibnamefont{and}
  \bibinfo{author}{\bibfnamefont{C.~P.} \bibnamefont{Herrero}},
  \bibinfo{journal}{Phys. Rev. B} \textbf{\bibinfo{volume}{72}},
  \bibinfo{pages}{024303} (\bibinfo{year}{2005}).

\bibitem[{\citenamefont{Herrero and Ram\'{\i}rez}(2010)}]{herrero10}
\bibinfo{author}{\bibfnamefont{C.~P.} \bibnamefont{Herrero}} \bibnamefont{and}
  \bibinfo{author}{\bibfnamefont{R.}~\bibnamefont{Ram\'{\i}rez}},
  \bibinfo{journal}{Phys. Rev. B} \textbf{\bibinfo{volume}{82}},
  \bibinfo{pages}{174117} (\bibinfo{year}{2010}).

\bibitem[{\citenamefont{Martyna et~al.}(1994)\citenamefont{Martyna, Tobias, and
  Klein}}]{martyna94}
\bibinfo{author}{\bibfnamefont{G.~J.} \bibnamefont{Martyna}},
  \bibinfo{author}{\bibfnamefont{D.~J.} \bibnamefont{Tobias}},
  \bibnamefont{and} \bibinfo{author}{\bibfnamefont{M.~L.} \bibnamefont{Klein}},
  \bibinfo{journal}{J. Chem. Phys.} \textbf{\bibinfo{volume}{101}},
  \bibinfo{pages}{4177} (\bibinfo{year}{1994}).

\bibitem[{\citenamefont{Ma et~al.}(2012)\citenamefont{Ma, Zheng, Sun, Yang, Xu,
  and Chu}}]{ma12}
\bibinfo{author}{\bibfnamefont{F.}~\bibnamefont{Ma}},
  \bibinfo{author}{\bibfnamefont{H.~B.} \bibnamefont{Zheng}},
  \bibinfo{author}{\bibfnamefont{Y.~J.} \bibnamefont{Sun}},
  \bibinfo{author}{\bibfnamefont{D.}~\bibnamefont{Yang}},
  \bibinfo{author}{\bibfnamefont{K.~W.} \bibnamefont{Xu}}, \bibnamefont{and}
  \bibinfo{author}{\bibfnamefont{P.~K.} \bibnamefont{Chu}},
  \bibinfo{journal}{Appl. Phys. Lett.} \textbf{\bibinfo{volume}{101}},
  \bibinfo{pages}{111904} (\bibinfo{year}{2012}).

\bibitem[{\citenamefont{Meyer et~al.}(2007)\citenamefont{Meyer, Geim,
  Katsnelson, Novoselov, Booth, and Roth}}]{meyer06}
\bibinfo{author}{\bibfnamefont{J.~C.} \bibnamefont{Meyer}},
  \bibinfo{author}{\bibfnamefont{A.~K.} \bibnamefont{Geim}},
  \bibinfo{author}{\bibfnamefont{M.~I.} \bibnamefont{Katsnelson}},
  \bibinfo{author}{\bibfnamefont{K.~S.} \bibnamefont{Novoselov}},
  \bibinfo{author}{\bibfnamefont{T.~J.} \bibnamefont{Booth}}, \bibnamefont{and}
  \bibinfo{author}{\bibfnamefont{S.}~\bibnamefont{Roth}},
  \bibinfo{journal}{Nature} \textbf{\bibinfo{volume}{446}}, \bibinfo{pages}{60}
  (\bibinfo{year}{2007}).

\bibitem[{\citenamefont{Adamyan et~al.}(2016)\citenamefont{Adamyan, Bondarev,
  and Zavalniuk}}]{adamyan16}
\bibinfo{author}{\bibfnamefont{V.}~\bibnamefont{Adamyan}},
  \bibinfo{author}{\bibfnamefont{V.}~\bibnamefont{Bondarev}}, \bibnamefont{and}
  \bibinfo{author}{\bibfnamefont{V.}~\bibnamefont{Zavalniuk}},
  \bibinfo{journal}{Physics Letters A} \textbf{\bibinfo{volume}{380}},
  \bibinfo{pages}{3732 } (\bibinfo{year}{2016}).

\bibitem[{\citenamefont{Bondarev et~al.}(2018)\citenamefont{Bondarev, Adamyan,
  and Zavalniuk}}]{bondarev18}
\bibinfo{author}{\bibfnamefont{V.~N.} \bibnamefont{Bondarev}},
  \bibinfo{author}{\bibfnamefont{V.~M.} \bibnamefont{Adamyan}},
  \bibnamefont{and} \bibinfo{author}{\bibfnamefont{V.~V.}
  \bibnamefont{Zavalniuk}}, \bibinfo{journal}{Phys. Rev. B}
  \textbf{\bibinfo{volume}{97}}, \bibinfo{pages}{035426}
  (\bibinfo{year}{2018}).

\bibitem[{\citenamefont{Amorim et~al.}(2014)\citenamefont{Amorim, Rold\'an,
  Cappelluti, Fasolino, Guinea, and Katsnelson}}]{amorim14}
\bibinfo{author}{\bibfnamefont{B.}~\bibnamefont{Amorim}},
  \bibinfo{author}{\bibfnamefont{R.}~\bibnamefont{Rold\'an}},
  \bibinfo{author}{\bibfnamefont{E.}~\bibnamefont{Cappelluti}},
  \bibinfo{author}{\bibfnamefont{A.}~\bibnamefont{Fasolino}},
  \bibinfo{author}{\bibfnamefont{F.}~\bibnamefont{Guinea}}, \bibnamefont{and}
  \bibinfo{author}{\bibfnamefont{M.~I.} \bibnamefont{Katsnelson}},
  \bibinfo{journal}{Phys. Rev. B} \textbf{\bibinfo{volume}{89}},
  \bibinfo{pages}{224307} (\bibinfo{year}{2014}).

\bibitem[{\citenamefont{Michel et~al.}(2015)\citenamefont{Michel, Costamagna,
  and Peeters}}]{michel15}
\bibinfo{author}{\bibfnamefont{K.~H.} \bibnamefont{Michel}},
  \bibinfo{author}{\bibfnamefont{S.}~\bibnamefont{Costamagna}},
  \bibnamefont{and} \bibinfo{author}{\bibfnamefont{F.~M.}
  \bibnamefont{Peeters}}, \bibinfo{journal}{physica status solidi (b)}
  \textbf{\bibinfo{volume}{252}}, \bibinfo{pages}{2433} (\bibinfo{year}{2015}).

\bibitem[{\citenamefont{Burmistrov et~al.}(2016)\citenamefont{Burmistrov,
  Gornyi, Kachorovskii, Katsnelson, and Mirlin}}]{burmistrov16}
\bibinfo{author}{\bibfnamefont{I.~S.} \bibnamefont{Burmistrov}},
  \bibinfo{author}{\bibfnamefont{I.~V.} \bibnamefont{Gornyi}},
  \bibinfo{author}{\bibfnamefont{V.~Y.} \bibnamefont{Kachorovskii}},
  \bibinfo{author}{\bibfnamefont{M.~I.} \bibnamefont{Katsnelson}},
  \bibnamefont{and} \bibinfo{author}{\bibfnamefont{A.~D.}
  \bibnamefont{Mirlin}}, \bibinfo{journal}{Phys. Rev. B}
  \textbf{\bibinfo{volume}{94}}, \bibinfo{pages}{195430}
  (\bibinfo{year}{2016}).

\bibitem[{\citenamefont{Ha\ifmmode~\check{s}\else \v{s}\fi{}\'{\i}k
  et~al.}(2018)\citenamefont{Ha\ifmmode~\check{s}\else \v{s}\fi{}\'{\i}k,
  Tosatti, and Marto\ifmmode~\check{n}\else \v{n}\fi{}\'ak}}]{hasik18}
\bibinfo{author}{\bibfnamefont{J.}~\bibnamefont{Ha\ifmmode~\check{s}\else
  \v{s}\fi{}\'{\i}k}},
  \bibinfo{author}{\bibfnamefont{E.}~\bibnamefont{Tosatti}}, \bibnamefont{and}
  \bibinfo{author}{\bibfnamefont{R.}~\bibnamefont{Marto\ifmmode~\check{n}\else
  \v{n}\fi{}\'ak}}, \bibinfo{journal}{Phys. Rev. B}
  \textbf{\bibinfo{volume}{97}}, \bibinfo{pages}{140301}
  (\bibinfo{year}{2018}).

\bibitem[{\citenamefont{Nicholl et~al.}(2015)\citenamefont{Nicholl, Conley,
  Lavrik, Vlassiouk, Puzyrev, Sreenivas, Pantelides, and Bolotin}}]{nicholl15}
\bibinfo{author}{\bibfnamefont{R.~J.~T.} \bibnamefont{Nicholl}},
  \bibinfo{author}{\bibfnamefont{H.~J.} \bibnamefont{Conley}},
  \bibinfo{author}{\bibfnamefont{N.~V.} \bibnamefont{Lavrik}},
  \bibinfo{author}{\bibfnamefont{I.}~\bibnamefont{Vlassiouk}},
  \bibinfo{author}{\bibfnamefont{Y.~S.} \bibnamefont{Puzyrev}},
  \bibinfo{author}{\bibfnamefont{V.~P.} \bibnamefont{Sreenivas}},
  \bibinfo{author}{\bibfnamefont{S.~T.} \bibnamefont{Pantelides}},
  \bibnamefont{and} \bibinfo{author}{\bibfnamefont{K.~I.}
  \bibnamefont{Bolotin}}, \bibinfo{journal}{Nature Comm.}
  \textbf{\bibinfo{volume}{6}}, \bibinfo{pages}{8789} (\bibinfo{year}{2015}).

\bibitem[{\citenamefont{Doussal and Radzihovsky}(2017)}]{doussal17}
\bibinfo{author}{\bibfnamefont{P.~L.} \bibnamefont{Doussal}} \bibnamefont{and}
  \bibinfo{author}{\bibfnamefont{L.}~\bibnamefont{Radzihovsky}},
  \bibinfo{journal}{Annals of Physics}  (\bibinfo{year}{2017}), ISSN
  \bibinfo{issn}{0003-4916},
  \urlprefix\url{http://www.sciencedirect.com/science/article/pii/S0003491617302531}.

\bibitem[{\citenamefont{Politano et~al.}(2012)\citenamefont{Politano, Marino,
  Campi, Far\'ias, Miranda, and Chiarello}}]{politano12}
\bibinfo{author}{\bibfnamefont{A.}~\bibnamefont{Politano}},
  \bibinfo{author}{\bibfnamefont{A.~R.} \bibnamefont{Marino}},
  \bibinfo{author}{\bibfnamefont{D.}~\bibnamefont{Campi}},
  \bibinfo{author}{\bibfnamefont{D.}~\bibnamefont{Far\'ias}},
  \bibinfo{author}{\bibfnamefont{R.}~\bibnamefont{Miranda}}, \bibnamefont{and}
  \bibinfo{author}{\bibfnamefont{G.}~\bibnamefont{Chiarello}},
  \bibinfo{journal}{Carbon} \textbf{\bibinfo{volume}{50}}, \bibinfo{pages}{4903
  } (\bibinfo{year}{2012}).

\bibitem[{\citenamefont{Jiang et~al.}(2015)\citenamefont{Jiang, Wang, Wang, and
  Park}}]{jiang_15}
\bibinfo{author}{\bibfnamefont{J.-W.} \bibnamefont{Jiang}},
  \bibinfo{author}{\bibfnamefont{B.-S.} \bibnamefont{Wang}},
  \bibinfo{author}{\bibfnamefont{J.-S.} \bibnamefont{Wang}}, \bibnamefont{and}
  \bibinfo{author}{\bibfnamefont{H.~S.} \bibnamefont{Park}},
  \bibinfo{journal}{J. Phys.: Condens. Matter} \textbf{\bibinfo{volume}{27}},
  \bibinfo{pages}{083001} (\bibinfo{year}{2015}).

\bibitem[{\citenamefont{Kumar et~al.}(2010)\citenamefont{Kumar, Hembram, and
  Waghmare}}]{kumar10}
\bibinfo{author}{\bibfnamefont{S.}~\bibnamefont{Kumar}},
  \bibinfo{author}{\bibfnamefont{K.~P. S.~S.} \bibnamefont{Hembram}},
  \bibnamefont{and} \bibinfo{author}{\bibfnamefont{U.~V.}
  \bibnamefont{Waghmare}}, \bibinfo{journal}{Phys. Rev. B}
  \textbf{\bibinfo{volume}{82}}, \bibinfo{pages}{115411}
  (\bibinfo{year}{2010}).

\bibitem[{\citenamefont{Waheed and Edholm}(2009)}]{waheed09}
\bibinfo{author}{\bibfnamefont{Q.}~\bibnamefont{Waheed}} \bibnamefont{and}
  \bibinfo{author}{\bibfnamefont{O.}~\bibnamefont{Edholm}},
  \bibinfo{journal}{Biophys. J.} \textbf{\bibinfo{volume}{97}},
  \bibinfo{pages}{2754} (\bibinfo{year}{2009}).

\bibitem[{\citenamefont{Ram\'{\i}rez and Herrero}(2018)}]{ramirez18}
\bibinfo{author}{\bibfnamefont{R.}~\bibnamefont{Ram\'{\i}rez}}
  \bibnamefont{and} \bibinfo{author}{\bibfnamefont{C.~P.}
  \bibnamefont{Herrero}} (\bibinfo{year}{2018}), \bibinfo{note}{to be
  published}.

\bibitem[{\citenamefont{Behroozi}(1996)}]{behrrozi_96}
\bibinfo{author}{\bibfnamefont{F.}~\bibnamefont{Behroozi}},
  \bibinfo{journal}{Langmuir} \textbf{\bibinfo{volume}{12}},
  \bibinfo{pages}{2289} (\bibinfo{year}{1996}).

\bibitem[{\citenamefont{Kirilenko et~al.}(2011)\citenamefont{Kirilenko,
  Dideykin, and Van~Tendeloo}}]{kirilenko11}
\bibinfo{author}{\bibfnamefont{D.~A.} \bibnamefont{Kirilenko}},
  \bibinfo{author}{\bibfnamefont{A.~T.} \bibnamefont{Dideykin}},
  \bibnamefont{and}
  \bibinfo{author}{\bibfnamefont{G.}~\bibnamefont{Van~Tendeloo}},
  \bibinfo{journal}{Phys. Rev. B} \textbf{\bibinfo{volume}{84}},
  \bibinfo{pages}{235417} (\bibinfo{year}{2011}).

\end{thebibliography}

\end{document}